\pdfoutput=1
\documentclass[PAPER, atlasdraft=false, cernpreprint=true, UKenglish, texmf, orcidlogo]{atlasdoc}
\usepackage{atlaspackage}
\usepackage{atlasbiblatex}

\usepackage{atlasphysics}
\ifthenelse{\boolean{AtlasDraft}}{\usepackage[output=true, shift=true]{}}{}

\graphicspath{{logos/}{figures/}}

\usepackage{ANA-HIGP-2024-34-PAPER-defs}
\usepackage{comment}
\usepackage{threeparttable}

\newcommand{\AtlasCoordFootnote}{%
ATLAS uses a right-handed coordinate system with its origin at the nominal interaction point (IP)
in the centre of the detector and the \(z\)-axis along the beam pipe.
The \(x\)-axis points from the IP to the centre of the LHC ring,
and the \(y\)-axis points upwards.
Polar coordinates \((r,\phi)\) are used in the transverse plane,
\(\phi\) being the azimuthal angle around the \(z\)-axis.
The pseudorapidity is defined in terms of the polar angle \(\theta\) as \(\eta = -\ln \tan(\theta/2)\) and is equal to the rapidity
$ y = \frac{1}{2} \ln \left( \frac{E + p_z}{E - p_z} \right) $ in the relativistic limit.
Angular distance is measured in units of \(\Delta R \equiv \sqrt{(\Delta y)^{2} + (\Delta\phi)^{2}}\).}

\AtlasTitle{Search for a resonance decaying into a scalar particle and a Higgs boson in the final state with two bottom quarks and two photons with 199 fb$^{-1}$ of data collected at $\sqrt{s}$=13 and 13.6~TeV with the ATLAS detector}

\AtlasAbstract{%
A search for the resonant production of a heavy scalar $X$ decaying into a lighter scalar $S$ and a Higgs boson, through the process $X \rightarrow S (\rightarrow b \bar{b})H ( \rightarrow \gamma\gamma)$, where the two photons are consistent with the Higgs boson decay, is performed. The search is conducted using integrated luminosities of 140 and 59~fb$^{-1}$ of proton--proton collision data at centre-of-mass energies of 13 and 13.6~TeV, respectively, recorded with the ATLAS detector at the LHC. The search is performed over the mass ranges of \mbox{170 $\leq$ $m_{X}$ $\leq$ 1000~GeV} and \mbox{15 $\leq$ $m_{S}$ $\leq$ 500~GeV}. %
No significant excess over the Standard Model background predictions is observed and limits at 95\% confidence level are set on the product of cross section and branching fraction for the process $X \rightarrow S (\rightarrow b \bar{b})H ( \rightarrow \gamma\gamma)$ at 13~TeV, ranging from 9 fb to 0.06 fb.
}
\AtlasRefCode{HIGP-2024-34}
\PreprintIdNumber{CERN-EP-2025-204}

\AtlasDate{\today}

\AtlasJournal{Phys.\ Lett.\ B.}
\AtlasJournalRef{\PLB 877 (2026) 140425}
\AtlasDOI{10.1016/j.physletb.2026.140425}

\AtlasCoverEgroupAnalysisTeam{atlas-HIGP-2024-34-analysis-team@cern.ch}

\hypersetup{pdftitle={ATLAS document},pdfauthor={The ATLAS Collaboration}}

\begin{document}

\maketitle

\section{Introduction} %
%

%
Measurements of the properties of the Higgs boson ($\hsm$), discovered in 2012 by the ATLAS and CMS Collaborations at the LHC~\cite{HIGG-2012-27,CMS-HIG-12-028}, show good agreement with the Standard Model (SM) predictions.
However, the current experimental results still allow for the presence of additional scalar states beyond the SM\@.
Such states are predicted in various models with an extended Higgs sector.
These include models with a complex singlet~\cite{PhysRevD.79.015018} or two real singlets~\cite{TRSM,Papaefstathiou_2021}, as well as the complex two-Higgs-doublet model (2HDM)~\cite{ginzburg2002twohiggsdoublet}, the 2HDM extended by a real scalar singlet (2HDM+\S)~\cite{PhysRevD79023521,Muhlleitner2017}  or the next-to-minimal supersymmetric Standard Model~\cite{ELLWANGER20101,ManiatisNMSSM}.  %
If two additional scalar bosons, $\X$ and $\S$, are introduced and the condition $\mX>\mS+\mHiggs$ is fulfilled, the decay $\X \to \S \hsm$ is kinematically allowed, leading to distinctive experimental signatures characteristic of such models.

The $\X \to \S \hsm$ process has been studied experimentally in various final states.
The CMS Collaboration has performed searches for $\X\to \Sbb \Hbb$~\cite{CMS-B2G-21-003}, \mbox{$\X\to \Sbb \Htautau$~\cite{CMS-HIG-20-014}} and $\X\to \Sbb \Hyy$~\cite{CMS-HIG-21-011}, as well as a statistical combination of these channels~\cite{CMS-B2G-23-002} using Run-2 data.
Similarly, ATLAS has explored this process in Run-2 data, with searches for $\X\to \SVV \Htautau$ and $\X\to \SVV \Hyy$~\cite{HDBS-2021-23}, where $V$ can be a $W^\pm$ or $Z$ boson~\cite{HDBS-2022-44}.
A related search targeting the $6b$ final state has also been performed in the context of triple Higgs boson production, interpreting the results in terms of the cascade decay $\X\to SH\to HHH$ with Run-2 data~\cite{HIGP-2024-32}.
Finally, a search for $\X\to \Sbb \Hyy$~\cite{HDBS-2021-17}, covering the mass ranges $170 \leq \mX \leq 1000$~\GeV and $15 \leq \mS \leq 500$~\GeV, has also been published by ATLAS using Run-2 data.
No significant excess was observed above the expected background and 95\% confidence level (CL) upper limits were set on the product of cross section and branching fraction for the process $X \rightarrow S (\rightarrow b \bar{b})H ( \rightarrow \gamma\gamma)$, ranging from 39~fb to 0.09~fb.
The largest deviation from the background-only hypothesis in this ATLAS search was observed at $(\mX, \mS)=(575, 200)$~\GeV with a local (global) significance of 3.5 (2.0) standard deviations.
In the same final state, CMS reported a deviation from the background-only hypothesis at $(\mX, \mS)=(650, 90)$~\GeV with a local (global) significance of 3.8 (<2.8) standard deviations~\cite{CMS-HIG-21-011}, which was not confirmed by the ATLAS analysis~\cite{HDBS-2021-17}.
More recently, CMS performed a search for $\X\to \Hbb \Syy$, and reported a deviation from the background-only hypothesis at $(\mX, \mS)=(300, 77)$~\GeV
with a local (global) significance of 3.3 (0.65) standard deviations~\cite{CMS-B2G-24-001}.

\begin{figure}[tbh]
\centering
\subfloat{\includegraphics[width=0.45\textwidth]{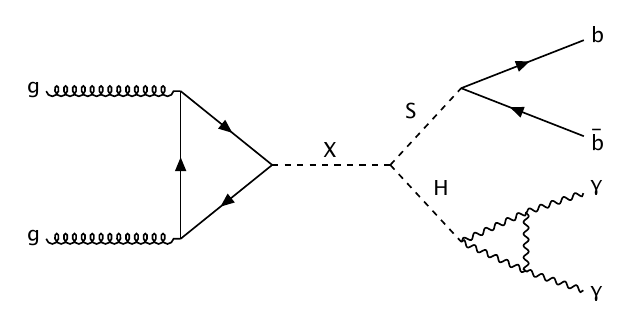}}
\caption{Illustrative Feynman diagram for $\X\to \Sbb \Hyy$ production via gluon--gluon fusion.}
\label{fig:feynman}
\end{figure}

This paper presents an updated search for $\X\to \Sbb \Hyy$ combining proton--proton ($pp$) collision data at 13~\TeV and 13.6~\TeV, recorded with the ATLAS detector in 2015--2018 and 2022--2023, respectively. The search probes the same phase-space as the previous analysis, offering unique sensitivity in the low-mass region with respect to other decay channels. Relative to the previous analysis, the object reconstruction, identification and calibration were improved. The strategy is overall similar but the training of the discriminating variable was redone and the definition of the signal regions was improved. The natural widths of the new bosons are assumed to be small compared with the experimental resolution.
Figure~\ref{fig:feynman} shows the representative leading-order Feynman diagram for the production and decay process studied.

This paper is organised as follows. %
Section~\ref{sec:detector} provides a brief description of the ATLAS detector.
Data and simulated event samples are described in Section~\ref{sec:samples}.
The object reconstruction is outlined in Section~\ref{sec:object_preselection}, followed by the analysis strategy in Section~\ref{sec:analysis_strategy}.
Systematic uncertainties are discussed in Section~\ref{sec:systematics}.
The statistical model, background validation, and final results are presented in Section~\ref{sec:results}, and conclusions are summarised in Section~\ref{sec:conclusion}.


%

%
\section{ATLAS detector} %
\label{sec:detector}

%
%

The ATLAS experiment~\cite{PERF-2007-01,GENR-2019-02} at the LHC is a multipurpose particle detector
with a forward--backward symmetric cylindrical geometry and a near \(4\pi\) coverage in
solid angle.\footnote{\AtlasCoordFootnote}
It consists of an inner tracking detector surrounded by a thin superconducting solenoid
providing a \qty{2}{\tesla} axial magnetic field, electromagnetic and hadronic calorimeters, and a muon spectrometer.
The inner tracking detector covers the pseudorapidity range \(|\eta| < 2.5\).
It consists of silicon pixel, silicon microstrip, and transition radiation tracking detectors.
Lead/liquid-argon (LAr) sampling calorimeters provide electromagnetic (EM) energy measurements
with high granularity within the region \(|\eta|< 3.2\).
A steel/scintillator-tile hadronic calorimeter covers the central pseudorapidity range (\(|\eta| < 1.7\)).
The endcap and forward regions are instrumented with LAr calorimeters
for EM and hadronic energy measurements up to \(|\eta| = 4.9\).
The muon spectrometer surrounds the calorimeters and is based on
three large superconducting air-core toroidal magnets with eight coils each.
The field integral of the toroids ranges between \num{2.0} and \qty{6.0}{\tesla\metre}
across most of the detector.
The muon spectrometer includes a system of precision tracking chambers up to \(|\eta| = 2.7\) and fast detectors for triggering up to \(|\eta| = 2.4\).
The luminosity measurement mainly relies on the LUCID--2 detector which is located close to the beampipe.
A two-level trigger system was used to select events~\cite{TRIG-2016-01,TRIG-2022-01}.
The first-level trigger is implemented in hardware and used a subset of the detector information
to accept events at a rate close to \qty{100}{\kHz}.
This is followed by a software-based trigger that
reduced the accepted rate of complete events to \qty{1.25}{\kHz} and \qty{3}{\kHz} on average in Run~2 and Run~3, respectively,
depending on the data-taking conditions.
A software suite~\cite{SOFT-2022-02} is used in data simulation, in the reconstruction
and analysis of real and simulated data, in detector operations, and in the trigger and data acquisition
systems of the experiment.


%

%
\section{Data and simulated event samples} %
\label{sec:samples}

%

%
This search uses $pp$ collision data at the LHC, collected by the ATLAS experiment during the full Run~2 at a centre-of-mass energy of $\sqrt{s}$ = 13~\TeV and during the initial Run-3 period (2022--2023) at 13.6~\TeV.
Data quality requirements are applied following the criteria described in Ref.~\cite{DAPR-2018-01}.
The resulting integrated luminosity corresponds to \runtwoLumi\ with an uncertainty of 0.83\% \cite{DAPR-2021-01} for Run~2 and \runthreeLumi\ with an uncertainty of 2.0\% \cite{ATL-DAPR-PUB-2024-001} for Run~3.

Data are selected using diphoton triggers that require two reconstructed photon candidates with transverse energies (\ET) of at least 35~\GeV\ and 25~\GeV\ \cite{TRIG-2018-05, TRIG-2022-01}.
In 2015 and 2016, the triggers require both photons to satisfy the \textit{Loose} identification criterion defined in Ref.~\cite{EGAM-2018-01}, while the \textit{Medium} criterion is applied for 2017--2018 and 2022--2023 to account for the higher $pp$ interaction rate.

Samples of Monte Carlo (MC) simulated signal events ($\X \to \S \hsm \to \bbyy$) are generated at leading-order (LO) accuracy in QCD in the gluon--gluon fusion (\ggF) production mode with \PYTHIA[8.3]~\cite{Bierlich:2022pfr}, using parton distribution functions (PDFs) from the \NNPDF[2.3lo] set~\cite{Ball:2012cx}.
The parton shower is simulated using \PYTHIA[8.3] with the A14 set of tuned parameters (tune)~\cite{ATL-PHYS-PUB-2014-021}.
The Higgs boson is required to decay into a pair of photons, while the scalar \S\ is set to decay into two $b$-quarks.
The widths of the \X\ and \S\ bosons are fixed to 10~\MeV in the simulation, well below the experimental resolution. The results of this type of searches are valid for \X\ boson natural widths $\Gamma_{\X}/\mX$ up to 2\%~\cite{HDBS-2023-17}.
Signal samples are generated for 160 mass points spanning the ranges 170~$\leq$~\mX~$\leq$~1000~\GeV\ and 15~$\leq$~\mS~$\leq$~500~\GeV. The theory production cross sections are calculated at LO in perturbative QCD by \PYTHIA[8.3]. The cross-section increase between 13~\TeV\ and 13.6~\TeV\ is of the order of 10\% for \mX\ values below 400~\GeV\ and up to 15\% for \mbox{\mX\ = 1~\TeV}.
The main backgrounds in this search arise from the continuum non-resonant diphoton production and processes in which a Higgs boson decays into \yy.
Production mechanisms via \ggF\ and vector boson fusion (\VBF) are considered for both SM single- and double-Higgs boson production. For the single-Higgs boson production, associated production with a $W$ or $Z$ boson (\WH and \ZH), production in association with a top-quark pair (\ttH), and other smaller processes are also considered.
The non-resonant diphoton background is composed mainly of processes with two prompt photons, but also events where one or both photon candidates are misidentified jets, such as $\gamma$+jets and dijet events.
These three contributions are collectively referred to as \yy+jets.
Smaller background contributions from \ttyy\ production, where a top-quark pair is produced in association with a diphoton final state, and from \Zqqyy production are added to the \yy+jet component of the non-resonant background. The normalisation of the \yy+jets background is fully determined from data, while a \SHERPA[2.2.14]~\cite{Bothmann:2019yzt} MC sample with two prompt photons in the matrix element is used for the kinematic modelling. %
For background processes with a Higgs boson decaying into \yy, the cross sections are set to the most precise available theory calculations~\cite{deFlorian:2016spz}. %
Branching ratios of 0.227\% and 58.2\% are assumed for the Higgs boson decay into two photons and two $b$-quarks~\cite{deFlorian:2016spz,Djouadi:1997yw}, respectively.

The overall simulation strategy, including the overlay of inelastic $pp$ collisions (pile-up), follows that of the previous search~\cite{HDBS-2021-17} in Run~2. In Run~3 the pile-up events are generated from a mix of \EPOS[2.0.1.4]~\cite{Werner:2005jf} and \PYTHIA[8.308]~\cite{Bierlich:2022pfr}. The \EPOS events were generated with the \EPOS LHC tune~\cite{Pierog:2013ria} and the \PYTHIA events with the A3 tune~\cite{ATL-PHYS-PUB-2016-017} and the \NNPDF[2.3lo]~\cite{Ball:2012cx} set of PDFs.

All nominal samples used in this analysis, including signal and background processes, are summarised in Table~\ref{tab:mc_samples}.
For each sample, the generator set-up is listed along with the PDF set, parton shower model and the underlying-event set of tuned parameters.
The samples use \evtgen~\cite{Lange:2001uf} for the modelling of $b$- and $c$-hadron decays except for the MC samples generated with \SHERPA.
A full simulation of the ATLAS detector~\cite{SOFT-2010-01} based on \GEANT~\cite{Agostinelli:2002hh} is used to reproduce the detector response for all background processes involving a Higgs boson and the \ttyy\ process. On the other hand, signal samples and the \yy+jets and \Zqqyy samples are processed with the fast simulation~\atlfastthree~\cite{SIMU-2018-04}, which employs \GEANT except for a parameterisation of the calorimeter response.
An alternative \yy+jets MC sample generated with \MGNLO~\cite{Alwall:2014hca} is considered for the evaluation of systematic uncertainties. The PDF set from \NNPDF[3.0nlo]~~\cite{Ball:2014uwa} is used in the matrix element, while the AZNLO~\cite{STDM-2012-23}~(A14) tune is used for the parton showering in Run~2 (Run~3).

\begin{table}[ht]
\centering
\caption{Summary of the nominal signal and background samples. The generator used in the simulation, the PDF set, the showering model and the set of tuned parameters are also provided. When different from Run~2, settings used for the simulation of the Run-3 samples are indicated in parentheses.
\label{tab:mc_samples}}
\resizebox{\columnwidth}{!}{%
\begin{tabular}{ lllll }
\toprule
\textbf{Process}         & \textbf{Generator}                           & \textbf{PDF set}           & \textbf{Showering}    & \textbf{Tune} \\
\midrule
$\X \to \S \hsm$         & \PYTHIA[8.3]~\cite{Bierlich:2022pfr} & \NNPDF[2.3lo]~\cite{Ball:2012cx} & \PYTHIA[8.3]~\cite{Bierlich:2022pfr} & A14~\cite{ATL-PHYS-PUB-2014-021} \\
\midrule
$\gamma\gamma+$jets      & \SHERPA[2.2.14]~\cite{Bothmann:2019yzt} & \NNPDF[3.0nnlo]~\cite{Ball:2014uwa}  &   \SHERPA[2.2.14] & -- \\
$t\bar{t} \gamma \gamma$ & \MGNLO~\cite{Alwall:2014hca}  &  \NNPDF[2.3lo] & \PYTHIA[8.3] & A14 \\
$\Zqqyy\ $ & \SHERPA[2.2.16] & \NNPDF[3.0nnlo] &   \SHERPA[2.2.16] & -- \\
\midrule
ggF $H$          & \AtlasMC{Nnlops}~\cite{Hamilton:2012np,Campbell:2012am,Hamilton:2012rf,Bozzi:2005wk,deFlorian:2011xf} & PDF4LHC15nlo (\PDFforLHC[21])~\cite{Butterworth:2015oua}&   \PYTHIA[8.3]  &  AZNLO~\cite{STDM-2012-23}~(A14)      \\
VBF $H$          & \POWHEGBOX[v2]~\cite{Nason:2004rx, Frixione:2007vw, Alioli:2010xd, Nason:2009ai} &  PDF4LHC15nlo  (\PDFforLHC[21])     & \PYTHIA[8.3] & AZNLO~(A14)   \\
$WH$             & \POWHEGBOX[v2]~\cite{Mimasu:2015nqa, Luisoni:2013cuh}                            &  PDF4LHC15nlo  (\PDFforLHC[21])       & \PYTHIA[8.3] & AZNLO~(A14)     \\
$qq\to ZH$       & \POWHEGBOX[v2]                           &  PDF4LHC15nlo (\PDFforLHC[21])       & \PYTHIA[8.3] & AZNLO~(A14)     \\
$gg\to ZH$       & \POWHEGBOX[v2]                            & PDF4LHC15nlo (\PDFforLHC[21])         & \PYTHIA[8.3] & A14    \\
$t\bar{t}H$      & \POWHEGBOX[v2]~\cite{Hartanto:2015uka}                                           &  PDF4LHC15nlo  (\PDFforLHC[21])  & \PYTHIA[8.3] & A14  \\
$b\bar{b}H$      & \POWHEGBOX[v2]~\cite{Campbell:2012am}                                            & \NNPDF[3.0nlo]\, (\PDFforLHC[21]) & \PYTHIA[8.3] & A14   \\
$tHq$            & \MGNLO                                                                           & \NNPDF[3.0nlo]  & \PYTHIA[8.3] & A14    \\
$tHW$            & \MGNLO                                                                           & \NNPDF[3.0nlo]  & \PYTHIA[8.3] & A14   \\
\midrule
ggF $HH$         & \POWHEGBOX[v2]~\cite{Nason:2009ai,powheg-sm,powheg-kappa-lambda}              & \PDFforLHC[21]          & \PYTHIA[8.3] & A14 \\
VBF $HH$         & \MGNLO                                                                           & \NNPDF[3.0nlo]  & \PYTHIA[8.3] & A14 \\
\bottomrule
\end{tabular}
}
\end{table}


%

%
\section{Object definitions} %
\label{sec:object_preselection}

%
%
%
%
%

Photon candidates are reconstructed from topological clusters of energy deposits in the EM calorimeter and calibrated as described in Ref.~\cite{EGAM-2018-01, EGAM-2021-02}, in the region $|\eta| < 2.37$, excluding the transition region between the barrel and endcap calorimeters $1.37 < |\eta| < 1.52$. %
The ATLAS detector simulation was slightly updated with respect to Ref.~\cite{EGAM-2018-01}, to take into account improvements in the understanding of the detector response, leading to additional in-situ scale and resolution corrections. The corresponding uncertainties are increased to reflect the Run~3 pile-up conditions and a change in optimal filtering coefficients for the LAr calorimeter readout.
To reduce the background from jets, photon candidates are required to fulfil the $Tight$ identification criteria based on shower shapes in the EM calorimeter and energy leakage into the hadronic calorimeter~\cite{EGAM-2021-01}. The hard-scatter vertex is selected as the vertex with the highest $\sum \pT^{2}$, where the sum is calculated using all tracks with a transverse momentum (\pT) higher than 500~\MeV, associated with that particular vertex. In the previous analysis~\cite{HDBS-2021-17}, the diphoton production vertex was defined using the two highest-\ET\ photons and additional information from the tracking systems, but the difference has a negligible impact on this analysis.
A calorimeter-based isolation transverse energy, defined as the sum of the transverse energies of positive-energy topological clusters within a cone of size $\Delta R = 0.2$ around the photon candidate, is required to be smaller than 0.065 times the transverse energy of the photon.
A track-based isolation, defined as the scalar sum of transverse momenta of tracks with $\pT > 1~\GeV$, satisfying loose quality criteria~\cite{EGAM-2019-01}, not associated with photon conversions, and originating from the vertex within a $\Delta R = 0.2$ cone around the photon candidate, is required to be less than 5\% of the photon transverse energy.

Jets are reconstructed using the anti-$k_t$ algorithm~\cite{Cacciari:2008gp, Fastjet} with a jet radius parameter $R = 0.4$. The inputs to this algorithm are particle-flow objects~\cite{PERF-2015-09}, which combine measurements from the inner detector and the calorimeters~\cite{PERF-2014-07} to improve the jet energy resolution and increase the jet reconstruction efficiency, especially at low jet \pT. Calibrations are applied to the jet energy scale (JES) and jet energy resolution (JER), which include components derived both from simulation and in situ measurements, documented thoroughly in Ref.~\cite{JETM-2018-05}. The associated systematic uncertainties are also evaluated using a series of simulation-based techniques and in-situ measurements, documented in Ref.~\cite{JETM-2018-05}. Jets from pile-up interactions are suppressed with the use of a jet-vertex-tagger~\cite{PERF-2014-03} and a forward jet-vertex-tagger~\cite{ATL-PHYS-PUB-2019-026}. All jets are required to have a \pT\ larger than 25~\GeV\ and an absolute rapidity lower than 4.4.

The flavour of jets is determined using a transformer neural network algorithm, GN2~\cite{FTAG-2023-05}. Contrary to the deep-learning neural network (DL1r) algorithm~\cite{FTAG-2019-07} used in the previous analysis, GN2 does not depend on any other low-level flavour tagging algorithms and a single training of the model optimises the different features of the algorithm. It demonstrates substantial improvements over the DL1r tagger~\cite{FTAG-2023-05}. The analysis makes use of a working point with a 85\% efficiency to select $b$-jets, and rejection factors of 90 for light-flavour jets and 5.9 for $c$-jets, as measured in simulated $\ttbar$ events. Compared with the DL1r working point with a 77\% $b$-jet efficiency, the $b$-tagging efficiency is improved by 8\% on average, for similar light-jet and $c$-jet rejections. Only jets in the central region ($|\eta| < 2.5$) are considered for $b$-tagging. Scale factors are applied to correct for differences in $b$-tagging efficiency between data and simulation.

The energy of $b$-tagged jets is corrected for the possible contribution of muons from semileptonic $b$-hadron decays, the contribution of undetected energy of neutrinos and out-of-cone effects~\cite{HIGP-2025-10}. Those corrections were rederived, leading to an increase of acceptance close to the \pT\ threshold.

Electrons are reconstructed and calibrated as described in Ref.~\cite{EGAM-2018-01}. They need to satisfy $|\eta| < 2.47$, excluding the calorimeter transition region, and $\pT > 10$~\GeV. The identification requirement applied uses a \textit{Medium} working point, while a \textit{Loose} isolation working point is used, with both these criteria described in Ref.~\cite{EGAM-2021-01}.
Muons are reconstructed as described in Ref.~\cite{MUON-2018-03}. They are required to have $|\eta| < 2.7$ and $\pT > 10$~\GeV. For identification the \textit{Medium} working point is selected, and a \textit{Loose} isolation working point, with both these criteria described in Ref.~\cite{MUON-2018-03}.
Electrons and muons are both matched to the primary vertex via requirements on the longitudinal and transverse impact parameters on the tracks, $|z_{0}|$ and $|d_{0}|$, respectively.

Overlap removal procedures are applied to avoid using the same detector signals to reconstruct multiple objects. In this analysis the priority is given to photons, by removing jets, electrons and muons within $\Delta R < 0.4$ of a selected photon. Next, jets within $\Delta R < 0.2$ of electrons are removed. Finally, electrons and muons within $\Delta R < 0.4$ of any remaining jet are removed.


%

%
\section{Analysis strategy} %
\label{sec:analysis_strategy}

%
The data analysis was performed using the analysis framework documented in \cite{atlas_easyjet}, developed within the ATLAS Collaboration for collision data analysis.
The general strategy is similar to that used in Ref.~\cite{HDBS-2021-17}. It is presented briefly here and the changes are highlighted. %

\subsection{Event selection}
\label{sec:object_preselection_preselection}

Events are selected if at least two photon candidates are found with an invariant mass \myy\ between 105 and 160~\GeV, and with transverse energies higher than $0.35\times\myy$ and $0.25\times\myy$ respectively, to target a SM Higgs boson decay. In order to reduce the \ttH\ background, events containing a muon or electron are rejected.
Only jets in the central region ($|\eta|<2.5$) are considered, requiring between two and five such jets, with exactly one or two $b$-tagged jets. %
Events with more than two $b$-tagged jets are discarded, as this was shown to have a negligible impact on the analysis sensitivity.%

A signal region (SR) is defined by selecting events with \myy\ around the Higgs boson mass peak of 125~\GeV.
Compared with the previous analysis~\cite{HDBS-2021-17} the \myy\ window is reduced from 120--130~\GeV\ to 122.5--127.5~\GeV.
This range corresponds to 1.1--1.7 times the full width at half maximum of the \myy\ distribution, depending on the mass of the $X$ particle. By narrowing the SR window the expected signal yield decreases by 10\% to 15\%, while the expected amount of background events is halved.
The control region (CR), defined as $\myy \in [105,122.5]\cup[127.5,160]\, \GeV$, is used to constrain the normalisation of the \yy+jets background.

Two exclusive categories of events are defined, based on the number of $b$-jets. Because of the large range of masses for the $X$ and $S$ particles, the kinematics can be very different. In particular, for $\mX \gg \mS + \mH$ two topologies are considered. The $S$ particle can be very boosted, so that its decay products are very collimated and the jets from two $b$-quarks can be reconstructed as a single $b$-jet. In another configuration the decay products of the $S$ particle can have a very low~\pT\ and one jet is not reconstructed. All signal events with $\mS < 0.09\times \mX$ (around 10\% of the probed mass points) are thus probed in the category with exactly one $b$-jet and the other ones are probed in the category with exactly two $b$-jets. For example, at $(\mX,\mS)=(250,\, 15)\,\GeV$, 98\% of the events with invariant masses compatible with the true masses are reconstructed in the 1 $b$-tagged category, while at $(\mX,\mS)=(575,\, 200)\,\GeV$, 100\% belong to the 2 $b$-tagged category, and at $(\mX,\mS)=(1000,\, 70)\,\GeV$, 70\% belong to the 1 $b$-tagged category.

\subsection{Multivariate analysis}

Multivariate discriminants are used to separate signal from background events in the SR defined in Section ~\ref{sec:object_preselection_preselection}. Parameterised neural networks (PNNs)~\cite{Baldi_2016} take as input a vector of event characteristics $\bar{x}$ and a vector of phase space parameters $\bar{\theta}$ and yield a response function that is parameterised in $\bar{\theta}$.
The parameterisation provides a unique discriminant for each signal hypothesis, separating the targeted signal events from background events. Therefore, for each value of $\bar{\theta} = (\mS, \mX)$, the PNN($\bar{\theta}$) is effectively a different observable.
The PNNs provide sensitivity over the considered mass range and allow interpolation to values of $\bar{\theta}$ not explicitly included in the training. They are parameterised in the plane of the two particle masses, $\bar{\theta} = (\mS, \mX)$. The binned distribution of the PNN score is used in the likelihood function as described in Section~\ref{sec:results}.

The PNNs are trained using all generated mass points for a given $b$-tagged category. For simplicity, only the largest background processes are included in the training using MC samples: $\gamma\gamma$+jets, \ttH, \ggF\ $H$ and \ZH\ for the 2 $b$-tagged category and $\gamma\gamma$+jets, \ttH, \ggF\ $H$, \ZH, \VBF\ $H$ and \HH\ for the 1 $b$-tagged category, as the remaining backgrounds are expected to have a negligible effect on the classifier. All events from the selection described in Section~\ref{sec:object_preselection_preselection} are used. Simulated events are divided into three folds such that the same events can be used for training, validation, and evaluation without overtraining. The PNNs are trained separately for the two $b$-tagged categories, as well as for the Run-2 and Run-3 datasets to account for the difference in centre-of-mass energy and pile-up. The PNN internal architectures are optimised using the same procedure as in the previous analysis~\cite{HDBS-2021-17}, using KerasTuner~\cite{chollet2015keras}, which chooses the hyper-parameters maximising the area under the receiver operating characteristic curve calculated on an evaluation set and using bayesian optimisation~\cite{garnett_bayesoptbook_2023}.

The most effective features for training the PNNs across the full phase space are the invariant masses of various combinations of final-state photons and $b$-tagged jets, and the transverse momentum of the $b$-tagged jets.
A modified definition of the invariant masses $\ensuremath{m_{bb\gamma\gamma}}$ (invariant mass of the diphoton and two $b$-jet system) and $\ensuremath{m_{b\gamma\gamma}}$ (invariant mass of the diphoton and $b$-jet system) is used: \mbox{$\ensuremath{m_{bb\gamma\gamma}^{*} = m_{bb\gamma\gamma} - (\myy - 125~\GeV)}$} and $\ensuremath{m_{b\gamma\gamma}^{*} = m_{b\gamma\gamma} - (\myy - 125~\GeV)}$.
It allows to remove the correlation between the PNN score and the \myy\ distribution, decreasing the uncertainty coming from the background normalisation that is detailed in Section~\ref{sec:background_estimation}.
For the 2 $b$-tagged category the input features are $\bar{x} = (\mbb, \mbbyystar)$ where \mbb\ is the invariant mass of the dijet system, and for the 1 $b$-tagged category the input variables are $\bar{x} = (\ptb, \mbyystar)$ where \ptb is the \pt of the $b$-tagged jet.
Including other training variables was tested but it did not improve the sensitivity significantly. In particular the \myy\ invariant mass is not used in the training because the expected improvement on the sensitivity is small and because it is used in the definition of the CR. %
\subsection{Background estimation}
\label{sec:background_estimation}

The shape of the different background components is taken from simulation. The processes containing one Higgs boson (\ggF\ $H$, \VBF\ $H$, \WH, \ZH, \ttH, \bbH, \tHq, \tHW) or two Higgs bosons (\ggF\ $HH$, \VBF\ $HH$) and the \ttyy\ and \Zqqyy\ processes are normalised using the state-of-the-art theory cross sections.
As previously described, the dominant non-resonant $\gamma\gamma$+jets background includes both prompt diphoton events, and $\gamma$+jets and dijet events where jets are misidentified as photons. After preselection the fraction of events with two prompt photons was measured for the \HHbbyy\ analysis~\cite{HIGP-2025-10}. This fraction is around 74\%, the rest being events with at least one jet misidentified as a photon. The fractions of these components in Run~2 and Run~3 are consistent within statistical uncertainties.
For the \yy+jets background, the \SHERPA\ MC simulation is used to evaluate the shape of the PNN distribution for each mass point.
This choice is motivated by previous studies~\cite{HDBS-2018-34}, which showed  that the shapes of the PNN distributions obtained from \yy+jets,  $\gamma$+jets and dijet samples are compatible within statistical uncertainties.
These studies also demonstrated that considering those background components, where one or both photon candidates are misidentified jets, has a negligible impact on the shape of the discriminant and the sensitivity of the analysis.
As such, the \yy+jets sample is used to model the full \yy+jets background for the shape estimation.
Normalisation factors for the \yy+jets process are extracted independently for Run~2 and Run~3 from the CRs by matching the total background prediction to data, and are computed separately for the 1 $b$-tagged and 2 $b$-tagged categories.
The resulting normalisation factors are $1.47 \pm 0.02$ (1 $b$-tagged) and $1.75 \pm 0.04$ (2 $b$-tagged) for Run~2, and $1.18 \pm 0.03$ (1 $b$-tagged) and $1.40 \pm 0.05$ (2 $b$-tagged) for Run~3, where the quoted uncertainties are statistical. These normalisation factors are used to correct the normalisation of the \yy+jets background components in the SRs and CRs before the likelihood fit described in Section~\ref{sec:results}. They are then allowed to float in the fit without any constraint. The difference between the two data-taking periods arises from the differences in the $b$-tagging and photon identification data-to-MC corrections.
After applying those factors, a residual discrepancy between the PNN distributions of the \yy+jets processes in data and simulation is still observed in the CR\@, barely covered by the theoretical uncertainties described in Section~\ref{sec:systematics}. It is mostly coming from discrepancies in the \mbbyystar and \mbyystar distributions, used as input to the PNN. An event-by-event correction is defined, based on a linear fit to the data/MC ratio as a function of \mbbyystar (\mbyystar) in the 2~$b$-tagged (1~$b$-tagged) category in the CR\@. This correction of the shape is applied to the \yy+jets background MC in both the SR and CR since the PNN and \myy\ distributions are mostly uncorrelated and the correction was shown to be independent from \myy. It was checked that it results in good agreement of the PNN score distributions with data, within uncertainties, in the CR. The normalisation factors are not affected by this correction on the shape.

Table~\ref{tab:prefit_yields} summarises the expected number of background events in each $b$-tagged category and data-taking period for the SRs and CRs. The uncertainties quoted in this table are the sum of statistical and systematic uncertainties, the latter being described in Section~\ref{sec:systematics}.

\begin{table*}[htb]
\begin{threeparttable}[b]
\scriptsize
\begin{center}
\caption{SR and CR yields for the for 1 $b$-tagged and 2 $b$-tagged categories, for the different processes before the likelihood fit described in Section~\ref{sec:results}. ``Other Single Higgs'' includes the following production modes: \VBF\ $H$, $WH$, $tHq$, and $tHW$.
}
\label{tab:prefit_yields}
\setlength{\tabcolsep}{5pt}
\begin{tabular}{l | r@{\(\,\pm\,\)}l r@{\(\,\pm\,\)}l | r@{\(\,\pm\,\)}l r@{\(\,\pm\,\)}l | r@{\(\,\pm\,\)}l r@{\(\,\pm\,\)}l |r@{\(\,\pm\,\)}l r@{\(\,\pm\,\)}l}
\toprule
& \multicolumn{8}{c|}{Run~2} &  \multicolumn{8}{c}{Run~3} \\
& \multicolumn{4}{c|}{1 $b$-tagged} &  \multicolumn{4}{c}{2 $b$-tagged}& \multicolumn{4}{c|}{1 $b$-tagged} &  \multicolumn{4}{c}{2 $b$-tagged} \\
& \multicolumn{2}{c}{\textbf{SR}}  & \multicolumn{2}{c|}{\textbf{CR}} & \multicolumn{2}{c}{\textbf{SR}} & \multicolumn{2}{c|}{\textbf{CR}} & \multicolumn{2}{c}{\textbf{SR}}  & \multicolumn{2}{c|}{\textbf{CR}} & \multicolumn{2}{c}{\textbf{SR}} & \multicolumn{2}{c}{\textbf{CR}} \\
\midrule
$\gamma\gamma$+jets\tnote{*}   & 2652 & 22 & 23740 & 200   & 259 & 4 & 2347 & 19  & 1100 & 22 & 9897 & 198  & 104.9 & 2.9 & 955 & 19 \\
$\ttyy$   & 3.84 & 0.20 & 34.2 & 1.8 & 2.78 & 0.18 & 25.2 & 1.6  & 1.80 & 0.12 & 16.0 & 1.1 & 1.33 & 0.13 & 11.8 & 1.2 \\
\Zqqyy   & 4.86 & 0.28 & 42.5 & 2.6 & 2.01 & 0.17 & 18.1 & 1.5 & 2.24 & 0.20 & 20.6 & 1.6 & 0.81 & 0.11 & 8.1 & 0.9 \\
\ggF $H$+\bbH   & 70 & 70 & 16 & 15 & 8 & 7 & 1.6 & 1.5  & 37 & 33 & 10 & 10 & 3.5 & 3.3 & 0.9 & 0.9 \\
\ttH   & 8.6 & 1.1 & 1.53 & 0.29 & 8.0 & 1.0 & 1.39 & 0.27 & 3.9 & 0.5 & 0.86 & 0.33 & 3.6 & 0.5 & 0.76 & 0.30 \\
\ZH   & 6.5 & 0.6 & 1.20 & 0.24 & 3.54 & 0.32 & 0.60 & 0.12 & 2.85 & 0.32 & 0.65 & 0.23 & 1.48 & 0.20 & 0.31 & 0.12 \\
Other Single Higgs   & 20 & 13 & 3.8 & 2.6  & 2.0 & 0.8 & 0.35 & 0.16  & 9 & 6 & 2.2 & 1.6 & 0.9 & 0.4 & 0.20 & 0.11 \\
$HH$   & 1.26 & 0.08 & 0.24 & 0.04 & 1.49 & 0.12 & 0.28 & 0.05 & 0.56 & 0.06 & 0.15 & 0.04 & 0.68 & 0.09 & 0.17 & 0.05 \\
\midrule
Total   & 2770 & 70 & 23840 & 200 & 286 & 8 & 2395 & 20 & 1158 & 41 & 9947 & 199 & 117 & 5 & 977 & 20 \\ \hline
\midrule
Data   & \multicolumn{2}{c}{2669} & \multicolumn{2}{c}{23838} & \multicolumn{2}{c}{287} & \multicolumn{2}{c}{2395} & \multicolumn{2}{c}{1158} & \multicolumn{2}{c}{9947} & \multicolumn{2}{c}{123} & \multicolumn{2}{c}{977} \\
\bottomrule
\end{tabular}
\begin{tablenotes}
\item[*] The normalisation factors for the \yy+jets background are applied, but their uncertainties are not included in the \yy+jets uncertainties.
\end{tablenotes}
\end{center}
\end{threeparttable}
\end{table*}

\subsection{Signal interpolation}
\label{sec:analysis_strategy_interpolation}

In order to be sensitive to the whole parameters space in the considered (\mX, \mS) plane ($[170,1000]~\GeV$, $[15,500]~\GeV$) a study was performed in Ref.~\cite{HDBS-2021-17} to determine the values of (\mX, \mS) to be probed, based on the resolution of the different invariant masses and results from signal injection tests. %
A detailed description of the signal interpolation procedure can be found in Ref.~\cite{HDBS-2021-17}.

While the set of 28 simulated signal samples in the 1 $b$-tagged category is sufficient, for the 2 $b$-tagged category the 132 simulated signal samples are not sufficient, and the shapes of the \mbbyystar\ and \mbb\ distributions have to be interpolated for 200 mass points.
The resulting grid, consisting of 360 mass points, is spaced finely enough to ensure that a signal located anywhere in the plane would produce a measurable excess at one of the tested grid points or its immediate neighbours.
To get the shape of the interpolated mass points, the four-vectors of the \X\ and \S\ particles of a high-statistics reference sample are recomputed in the rest frame of the interpolated \X\ particle using a Lorentz transformation defined by the four-vector of \X. %
The rest-frame four-vectors of the \X\ and \S\ particles of this sample are rescaled so that they are distributed around the theoretical values of the interpolated point, and its events are weighted to reproduce the experimental resolution at that mass. Only the resolution of the \mbb\ mass is taken into account since the resolution effects of jets are much larger than for photons. For that purpose, the \mbb\ resolution of each simulated point is modelled using a Bukin probability density function~\cite{Bukin} and the values of the Bukin parameters are then interpolated to the interpolated masses.
The yield for each interpolated mass point is computed from the yields of the simulated samples by using Delaunay triangulation~\cite{DelaunayInterP}.
%


%

%
%
%
%
%

%
\section{Systematic uncertainties} %
\label{sec:systematics}

%

Experimental and theory systematic uncertainties are considered both for signal and the different types of background processes. Even for the mass point with the largest systematic uncertainty, the results are dominated by the statistical uncertainty. The systematic uncertainties can affect the yield (for example the uncertainty in the luminosity or the Higgs boson branching ratios), the shape of the PNN score distribution (uncertainties related to $\gamma\gamma$+jets background), or both (for example the experimental uncertainties). These uncertainties are computed independently for Run~2 and Run~3. The uncertainties in the luminosity, normalisation factors for the \yy+jets process, flavour-tagging and all uncertainties related to photons (trigger, calibration, identification and isolation) are not correlated between the two data-taking periods, while all theory and other experimental uncertainties are treated as fully correlated between the two data-taking periods. The uncertainties are also computed separately for the 1 $b$-tagged and 2 $b$-tagged categories. Systematic uncertainties are pruned if their variation is smaller than 0.5\% in yield or, for shape systematics, in the maximal per-bin deviation after yield normalisation.

Experimental systematic uncertainties are related to the luminosity, diphoton trigger, the photon and jet energy scale and resolution, as well as the photon identification and the jet flavour tagging. They are applied to the signal and all background processes. Their impact on the signal and backgrounds yields are of the order of a few percent.

Theory systematic uncertainties account for missing higher-order corrections in perturbative QCD calculations, uncertainties related to the parton shower modelling and uncertainties in the cross-section predictions.
Uncertainties due to missing higher-order QCD terms are evaluated by varying the renormalisation and factorisation scales from 0.5 to 2, and using the envelope of variations relative to the nominal scale, avoiding variations in opposite directions. Additional theory uncertainties arise from the limited knowledge of the PDFs and of the strong coupling constant  $\alpha_{S}$. These are estimated following the prescription in Ref.~\cite{Butterworth:2015oua}. They are computed for the signal, the non-resonant \SHERPA\ $\gamma\gamma$+jets and \ttyy\ backgrounds, the Higgs boson backgrounds \ttH, \ZH, and for \ggF $HH$. A 100\% theory uncertainty is assigned to the normalisation of the \ggF \,\hsm, \VBF \,\hsm\ and \WH\ processes, due to the limited knowledge of the modelling of the radiation of additional heavy-flavour jets~\cite{TOPQ-2017-12, STDM-2012-11, TOPQ-2012-16, HIGG-2018-29, HIGG-2019-13}. For the smaller Higgs boson backgrounds the uncertainty in the total cross section~\cite{deFlorian:2016spz} is used.
Uncertainties in the Higgs boson branching ratios to pairs of $b$-quarks and photons are also taken into account, following the prescriptions of Ref.~\cite{deFlorian:2016spz}.

A modelling uncertainty is computed for the continuum $\gamma\gamma$+jets background, by comparing the PNN distributions evaluated with the nominal \SHERPA\ samples and the alternative \MGNLO\ sample. This uncertainty is computed separately for Run~2 and Run~3 and the $b$-tagged categories. This uncertainty is of the order of 10-20\% in the 1 $b$-tagged category, and of the order of a few tens of percent in the 2 $b$-tagged category, with a few values up to 80\%.
The normalisation of the $\gamma\gamma$+jets background is determined by matching the background prediction to data in the CR, so systematic uncertainties for this process only affect the PNN shape. Because the same normalisation factor is applied in both the CR and SR, an uncertainty is assigned to account for possible mismodelling in the extrapolation from CR to SR\@. This is computed as the sum of the theory uncertainties in the SR/CR ratio and amounts to 1.3\% (0.3\%) for the 2 $b$-tagged (1 $b$-tagged) categories.

A shape uncertainty associated with the event-by-event linear correction of the \yy+jets background PNN shape is applied, based on \mbbyystar (\mbyystar) in the 2 $b$-tagged (1 $b$-tagged) category. This shape uncertainty reflects the uncertainty affecting the linear fit parameters. It is of the order of a few percent for most of the mass range, up to 15\% for \mX\ around 1000~\GeV, and its impact on the results is found to be up to approximately 2\%.

In the case of the signal PNN distribution obtained by the interpolation procedure described in Section~\ref{sec:analysis_strategy_interpolation}, an additional uncertainty in the shape is computed, coming from varying the parameters of the Bukin probabilities by their uncertainties. A closure test is performed to assess a global uncertainty in the yields of the interpolated points: each simulated mass point is removed one by one from the grid used for the Delaunay triangulation and the resulting interpolated yield is compared with the actual yield. This results in uncertainties of 4\% and 5\%, computed separately for the Run-2 and Run-3 data-taking periods, respectively.

The impact of the systematic uncertainties on the limits, after the profile-likelihood fit described in the following section, depends on the category and the mass of the \X\ particle.
For \mX\ below 400~\GeV\ the dominant contributions come from the generator comparison and the missing higher-order uncertainties for the $\gamma\gamma$+jets process, and the 100\% uncertainty for the \ggF \,\hsm\ normalisation; their effect depends on the category. In the 1 $b$-tagged category, the resulting changes in the limits are of the order of 10\% or less. In the 2 $b$-tagged category, the effect on the limits ranges from about 10 to 40\% for \mX\ below 250~\GeV and is of the order of 10\% above.
For \mX\ above 400~\GeV\ the dominant effects originate from the signal parton shower and PDF+$\alpha_{S}$ uncertainties with impacts of the order of 10\% or less in both categories.


%

%
\section{Results} %
\label{sec:results}

%
%
%
%
%

%
%
%

The statistical procedure applied for extracting results is similar to that in Ref.~\cite{HDBS-2021-17}, with updates to incorporate both Run-2 and Run-3 data.
A simultaneous binned maximum-likelihood fit is performed on the PNN output distributions in each SR and on the single bin of its corresponding CR\@. Those fits are done separately for the 1 $b$-tagged and 2 $b$-tagged categories but simultaneously for the two data-taking periods.
The binning of the PNN score is optimised independently for each signal hypothesis and data-taking period.
The procedure begins with a very fine binning of the score distribution.
Starting from the high-score end, adjacent bins are iteratively merged until the rightmost bin contains at least one expected background event.
Subsequent bins are defined by continuing the merging process: if $i$ is the bin index relative to the rightmost bin and $n_i$ its event content, it is required that each bin to the left contains at least $n_i+1$ expected background events.
This process continues until the expected signal-to-background ratio in the next candidate bin falls below that of the full unbinned distribution.
All remaining bins in the low-score region are then merged into a single bin with a background-dominated composition.

The signal strength $\mu$ is treated as the parameter of interest and is defined as the ratio of the observed event rate to the expected event rate, where the latter is computed assuming a 13~\TeV cross section equal to the observed upper limits from the previous analysis~\cite{HDBS-2021-17}, and the 13.6~\TeV cross section as introduced in Section~\ref{sec:samples}.
The choice of the 13~\TeV cross-section value is arbitrary, as it does not impact the limits or observed significance values, while the ratio of cross sections between the two centre-of-mass energies is needed.
Systematic uncertainties, described in Section~\ref{sec:systematics}, are included as nuisance parameters in the fit.
The likelihood function is constructed as a product of Poisson terms over the bins of the PNN output in the signal and control regions, following the same structure as in Ref.~\cite{HDBS-2021-17}.
Upper limits on $\mu$ are derived using a profile-likelihood-ratio test statistic~\cite{Cowan:2010js} and the CL$_s$ method~\cite{Read:2002hq} with the asymptotic approximation used to determine the confidence intervals. The upper limits on the signal cross section at 13~\TeV, $\sigma_{\mathrm{13~\TeV}}$, are then derived by using the reference cross sections.
The binning in PNN score was chosen to ensure that the asymptotic approximation would be valid and it was checked with pseudo-experiment studies both in Ref.~\cite{HDBS-2021-17} and in the updated analysis combining Run-2 and Run-3 data, showing good agreement well within the one-standard-deviation band.

\begin{figure}[!htb]
\centering
\subfloat[]{\includegraphics[width=0.45\textwidth]{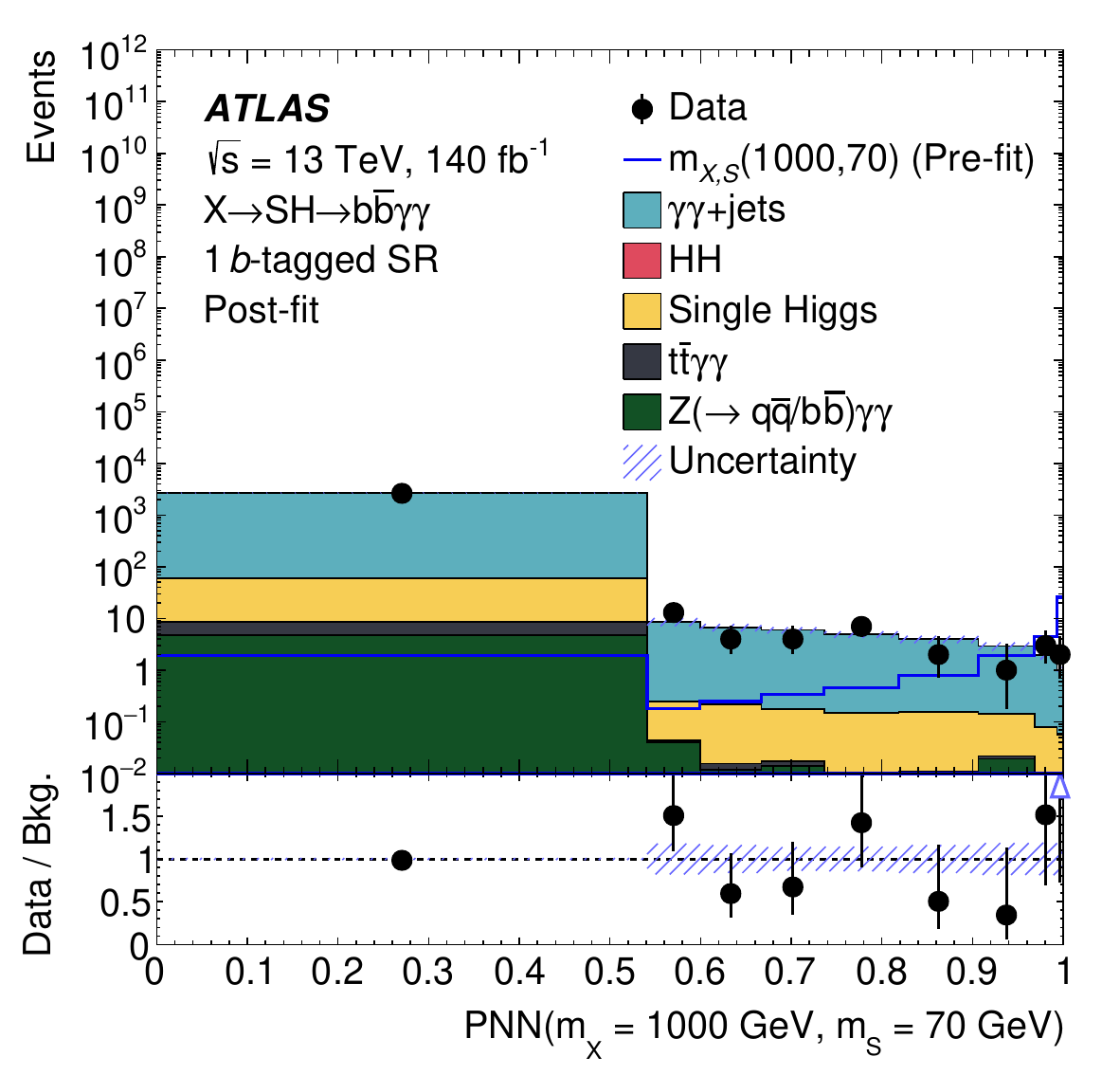}}
\subfloat[]{\includegraphics[width=0.45\textwidth]{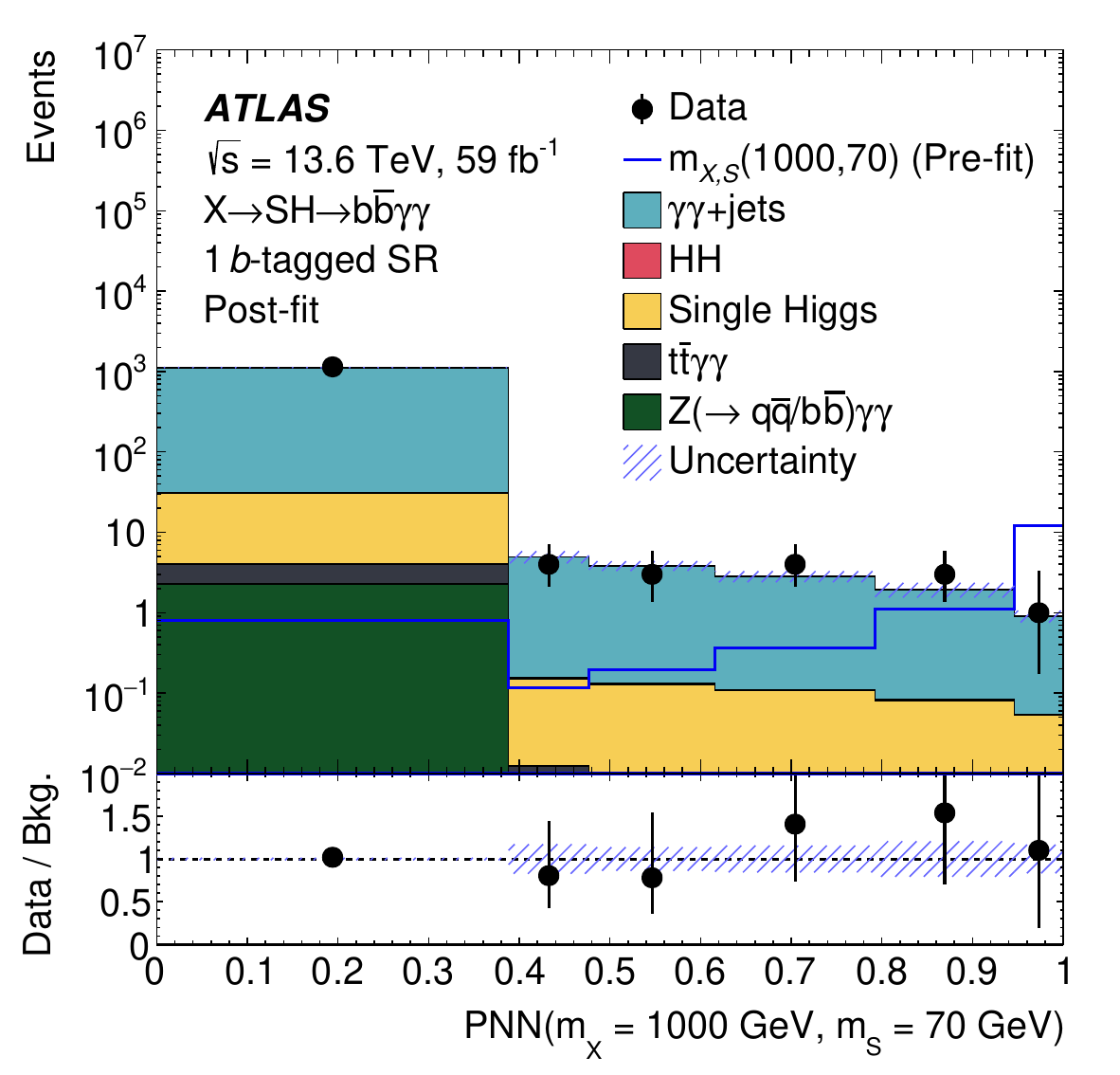}}\\
\subfloat[]{\includegraphics[width=0.45\textwidth]{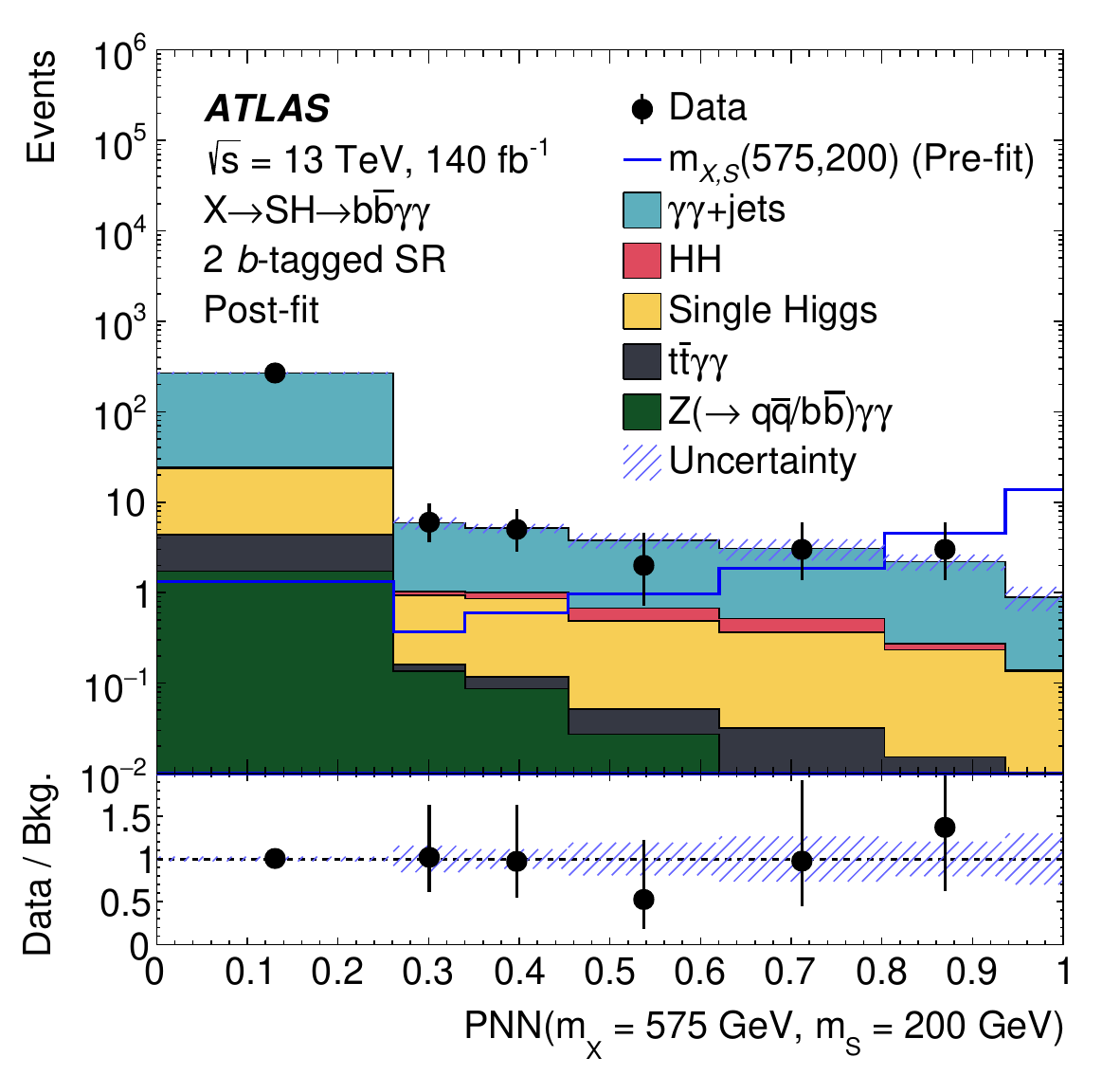}}
\subfloat[]{\includegraphics[width=0.45\textwidth]{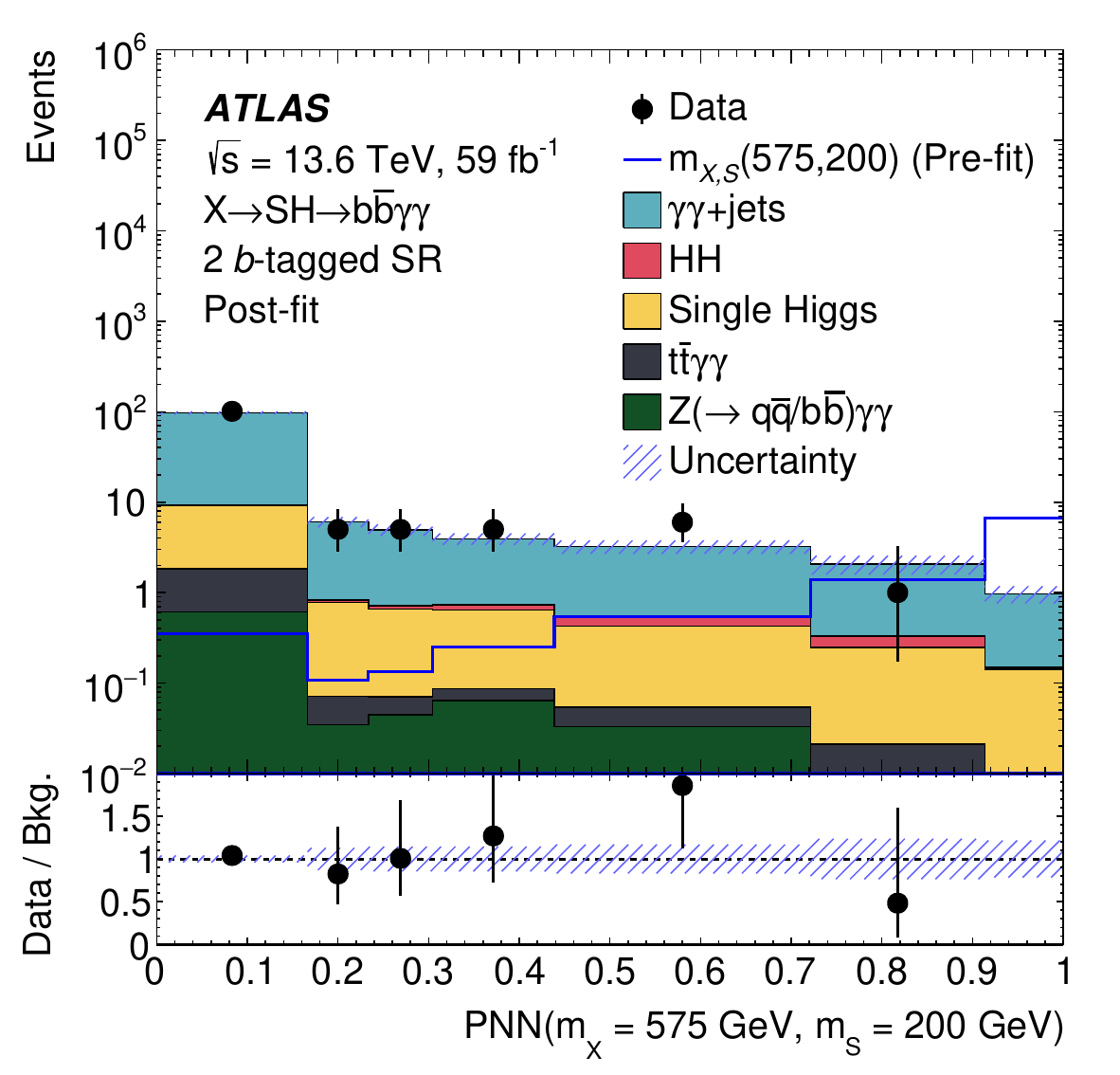}}
\caption{Distributions of the PNN discriminant output after the profile-likelihood fit in the SRs. Figures (a) and (b) show the 1 $b$-tagged SR for $\mX\ = 1000~\GeV$ and $\mS = 70~\GeV$ using (a) Run-2 and (b) Run-3 data. Figures (c) and (d) show the 2 $b$-tagged SR for $\mX\ = 575~\GeV$ and $\mS = 200~\GeV$ using (c) Run-2 and (d) Run-3 data. The \yy+jets category represents the sum of \yy+jets, $\gamma$+jets and dijet processes. The shaded band represents the total systematic uncertainty after the profile-likelihood fit. The blue line represents the signal normalised using an arbitrary cross section of 1 fb.}
\label{fig:PNN_fits_postfit}
\end{figure}

Figure~\ref{fig:PNN_fits_postfit} shows the PNN output distributions after the profile-likelihood fit for two signal mass points in the 1 $b$-tagged and 2 $b$-tagged categories. %
Overall, good agreement is observed between data and the SM background-only expectation for all probed mass points.

\begin{figure}[!htb]
\centering
\subfloat[]{\includegraphics[width=0.5\textwidth]{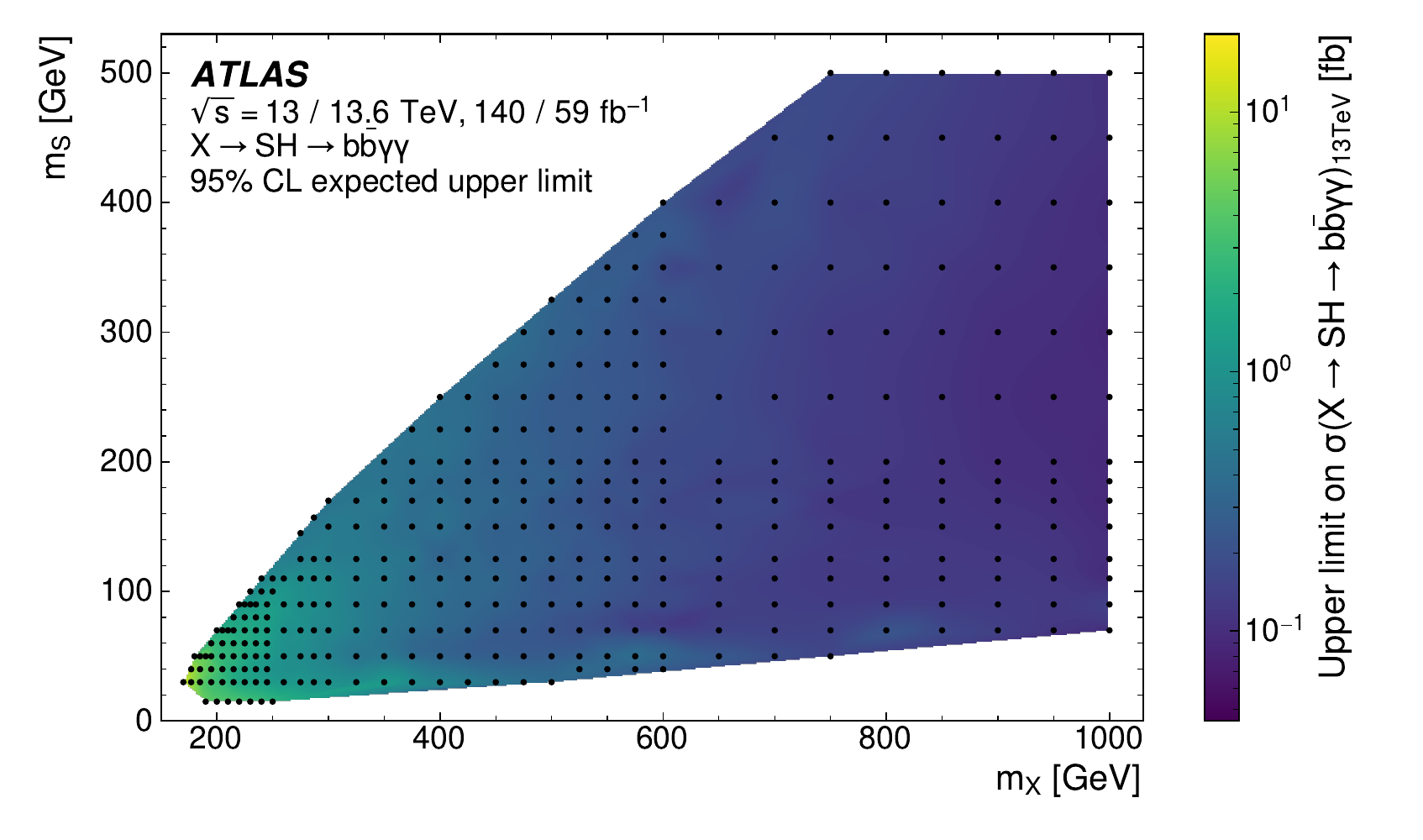}}
\subfloat[]{\includegraphics[width=0.5\textwidth]{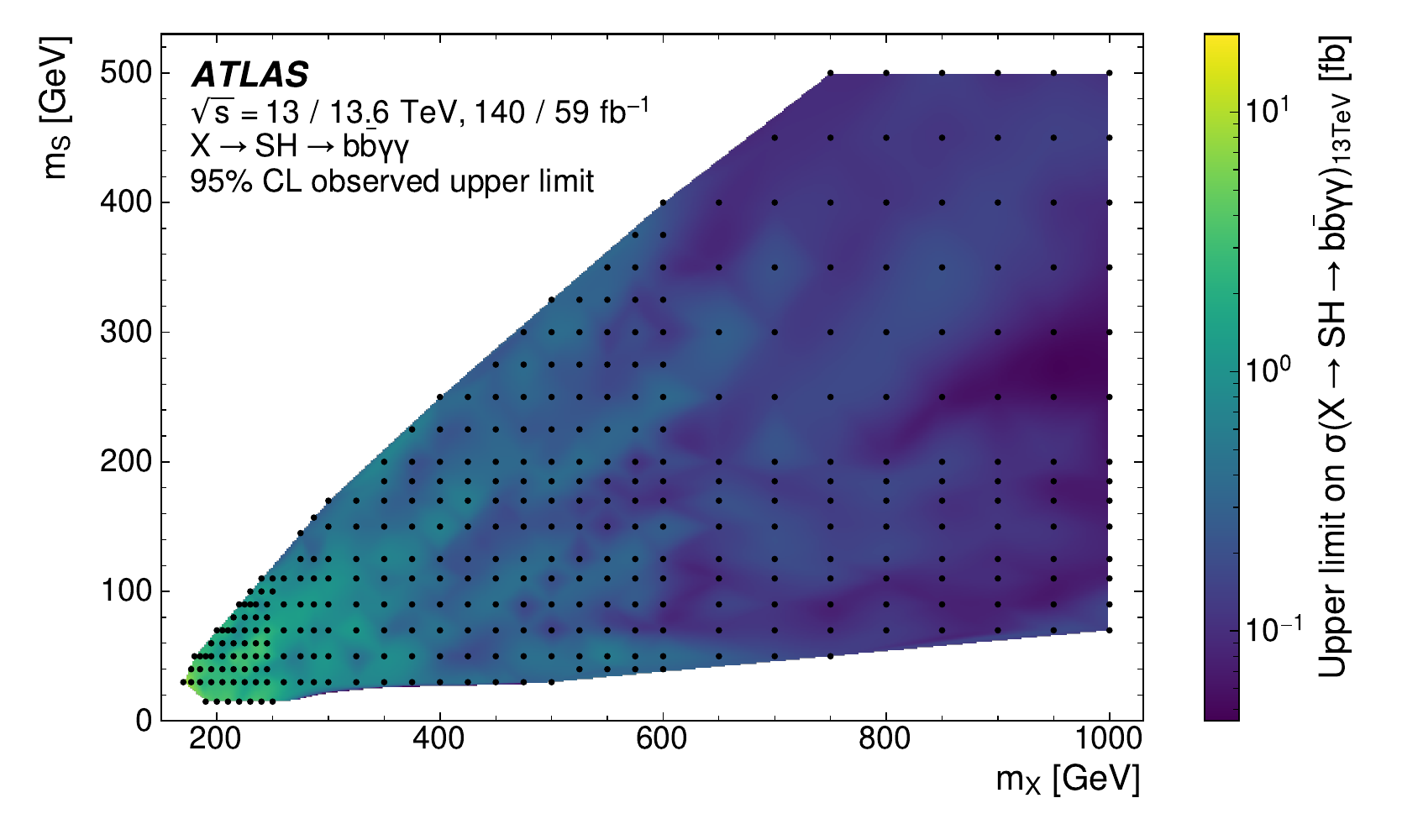}}
\caption{(a) Expected and (b) observed 95\% CL upper limits on the product of cross section and branching fraction for the process $\X\to \Sbb \Hyy$ at 13~\TeV, in the full (\mX, \mS) plane, including both the 1 $b$-tagged and 2 $b$-tagged categories.}
\label{fig:upper_limit13TeV_results}
\end{figure}

No significant excess is observed, the maximum observed local significance being 2.0 standard deviations, seen at $(\mX,\mS)=(235,\, 60)\,\GeV$.
Figure~\ref{fig:upper_limit13TeV_results} presents the expected and observed 95\% CL upper limits on $\sigma_{\mathrm{13~\TeV}}$ as a function of the (\mX, \mS) signal mass hypotheses, obtained from the combination of Run-2 and Run-3 data.
The observed (expected) limits range from 9.0 (11.4) fb at $\mX=170~\GeV$ and $\mS=30~\GeV$ to 0.06 (0.09) fb at  $\mX=1000~\GeV$ and \mS between 175 and 300 \GeV.
The analysis generally achieves better sensitivity in the high-mass region, where the signal kinematics differ more significantly from those of the background.
Compared with the previous Run-2 analysis~\cite{HDBS-2021-17}, the current search achieves significantly improved sensitivity across the entire (\mX, \mS) plane.
The expected limits on $\sigma_{\mathrm{13~\TeV}}$ are 15\% to 73\% better, depending on the mass, with the largest improvements in the low-mass region.
This improvement is driven by the inclusion of Run-3 data, a tighter \myy\ selection, the update of the $b$-jet energy correction, the adoption of the GN2 $b$-tagging algorithm with a higher efficiency (85\% instead of 77\%), and updates to the PNN training.
On the Run-2 data sample alone, the average improvement in the expected limit on $\sigma_{\mathrm{13~\TeV}}$ is about 20\%, going up to about 65\% at low-mass.
The reduction of the \myy\ window defining the SR brings an improvement of 10\% on average, while the use of the GN2 $b$-tagging algorithm with an efficiency working point of 85\% improves the expected limit by 10 to 65\% depending on the mass point.

In the previous ATLAS search~\cite{HDBS-2021-17} with Run-2 data only, the largest deviation from the background-only hypothesis was observed at $(\mX,\mS)=(575,\, 200)\,\GeV$, with a local (global) significance of 3.5 (2.0) standard deviations.
No similar deviation is observed at this mass point in the current search, neither in the Run-2 data sample alone nor in the combined Run-2 and Run-3 result (the $p$-value of the background-only hypothesis is larger than 0.5). The observed (expected) 95\% CL upper limit on $\sigma_{\mathrm{13~\TeV}}$ for this mass hypothesis is 0.12 (0.19) fb. The improvement on the expected 95\% CL upper limit relative to the previous search is of the order of 40\%.
The events contributing to the earlier excess of data were checked individually.
Most of them fail to pass the current 2 $b$-tagged preselection, either because a photon is now reconstructed as an electron, or because a jet is no longer tagged as $b$-jet. A small fraction of these events now also fall outside the \myy\ SR due to the narrower mass window used in the present analysis.
The excess of events seen by the CMS Collaboration using Run-2 data~\cite{CMS-HIG-21-011} was not observed in the previous ATLAS analysis~\cite{HDBS-2021-17} and is not observed in this analysis either (the $p$-value of the background-only hypothesis is 0.5).

%

%
%
\FloatBarrier

\section{Conclusion} %
\label{sec:conclusion}

A search for a hypothetical scalar boson \X\ decaying into another scalar \S\ and a SM Higgs boson $H$ is presented, targeting final states with two photons and two $b$-jets from the subsequent $\Htoyy$ and $\Stobb$ decays.
The analysis is performed in two signal regions based on the number of $b$-tagged jets, using parameterised neural networks trained to provide sensitivity across a wide range of mass hypotheses in the (\mX, \mS) plane.
The results are based on the full Run-2 (140~fb$^{-1}$) and early Run-3 (59~fb$^{-1}$) data samples collected by the ATLAS detector at the LHC\@.
Signal shape interpolation is used in the high-mass region to ensure smooth coverage between simulated grid points, while a finer grid of signal samples is employed at lower masses.

No significant deviation from the SM expectation is observed.
Upper limits at 95\% CL are set on the product of cross section and branching fraction for the process $X \rightarrow S (\rightarrow b \bar{b})H ( \rightarrow \gamma\gamma)$ at 13~\TeV for masses in the ranges \mbox{$170 \leq \mX \leq 1000~\GeV$} and \mbox{15~$\leq$~\mS~$\leq$~500~\GeV}.
This result extends the sensitivity of the previous analysis~\cite{HDBS-2021-17} through the inclusion of Run-3 data and several analysis improvements. The overall sensitivity of the analysis is improved by 15--73\%, depending on the mass, with the largest improvements in the low-mass region. The inclusion of the early Run-3 dataset contributes to improvements of the sensitivity of 9--30\%.


%
%
%
%
%
\section*{Acknowledgements}
%

%

%
%

%
%

We thank CERN for the very successful operation of the LHC and its injectors, as well as the support staff at
CERN and at our institutions worldwide without whom ATLAS could not be operated efficiently.

The crucial computing support from all WLCG partners is acknowledged gratefully, in particular from CERN, the ATLAS Tier-1 facilities at TRIUMF/SFU (Canada), NDGF (Denmark, Norway, Sweden), CC-IN2P3 (France), KIT/GridKA (Germany), INFN-CNAF (Italy), NL-T1 (Netherlands), PIC (Spain), RAL (UK) and BNL (USA), the Tier-2 facilities worldwide and large non-WLCG resource providers. Major contributors of computing resources are listed in Ref.~\cite{ATL-SOFT-PUB-2026-001}.

We gratefully acknowledge the support of ANPCyT, Argentina; YerPhI, Armenia; ARC, Australia; BMWFW and FWF, Austria; ANAS, Azerbaijan; CNPq and FAPESP, Brazil; NSERC, NRC and CFI, Canada; CERN; ANID, Chile; CAS, MOST and NSFC, China; Minciencias, Colombia; MEYS CR, Czech Republic; DNRF and DNSRC, Denmark; IN2P3-CNRS and CEA-DRF/IRFU, France; SRNSFG, Georgia; BMFTR, HGF and MPG, Germany; GSRI, Greece; RGC and Hong Kong SAR, China; ICHEP and Academy of Sciences and Humanities, Israel; INFN, Italy; MEXT and JSPS, Japan; CNRST, Morocco; NWO, Netherlands; RCN, Norway; MNiSW, Poland; FCT, Portugal; MNE/IFA, Romania; MSTDI, Serbia; MSSR, Slovakia; ARIS and MVZI, Slovenia; DSI/NRF, South Africa; MICIU/AEI, Spain; SRC and Wallenberg Foundation, Sweden; SERI, SNSF and Cantons of Bern and Geneva, Switzerland; NSTC, Taipei; TENMAK, T\"urkiye; STFC/UKRI, United Kingdom; DOE and NSF, United States of America.

Individual groups and members have received support from BCKDF, CANARIE, CRC and DRAC, Canada; CERN-CZ, FORTE and PRIMUS, Czech Republic; COST, ERC, ERDF, Horizon 2020, ICSC-NextGenerationEU and Marie Sk{\l}odowska-Curie Actions, European Union; Investissements d'Avenir Labex, Investissements d'Avenir Idex and ANR, France; DFG and AvH Foundation, Germany; Herakleitos, Thales and Aristeia programmes co-financed by EU-ESF and the Greek NSRF, Greece; BSF-NSF and MINERVA, Israel; NCN and NAWA, Poland; La Caixa Banking Foundation, CERCA Programme Generalitat de Catalunya and PROMETEO and GenT Programmes Generalitat Valenciana, Spain; G\"{o}ran Gustafssons Stiftelse, Sweden; The Royal Society and Leverhulme Trust, United Kingdom; United States of America.

In addition, individual members wish to acknowledge support from CERN: European Organization for Nuclear Research (CERN DOCT); Chile: Agencia Nacional de Investigaci\'on y Desarrollo (ANID FONDECYT reg. 1230987, FONDECYT 1230812, FONDECYT 1240864, Fondecyt 3240661, Fondecyt Regular 1240721); China: Chinese Ministry of Science and Technology (MOST-2023YFA1605700, MOST-2023YFA1609300), National Natural Science Foundation of China (NSFC - 12175119, NSFC 12275265); Czech Republic: Czech Science Foundation (GACR - 24-11373S), Ministry of Education Youth and Sports (ERC-CZ-LL2327, FORTE CZ.02.01.01/00/22\_008/0004632), PRIMUS Research Programme (PRIMUS/21/SCI/017); EU: H2020 European Research Council (ERC - 101002463); European Union: European Research Council (BARD No. 101116429, ERC - 948254, ERC 101089007), European Regional Development Fund (HE COFUND GA No.101081355, ERDF), Horizon 2020 Framework Programme (MUCCA - CHIST-ERA-19-XAI-00), European Union, Future Artificial Intelligence Research (FAIR-NextGenerationEU PE00000013), Italian Center for High Performance Computing, Big Data and Quantum Computing (ICSC, NextGenerationEU), Marie Sklodowska-Curie Actions (GAP-101168829); France: Agence Nationale de la Recherche (ANR-21-CE31-0013, ANR-21-CE31-0022, ANR-22-EDIR-0002, ANR-24-CE31-0504-01); Germany: Baden-Württemberg Stiftung (BW Stiftung-Postdoc Eliteprogramme), Deutsche Forschungsgemeinschaft (DFG - 469666862, DFG - CR 312/5-2); China: Research Grants Council (GRF); Italy: Istituto Nazionale di Fisica Nucleare (ICSC, NextGenerationEU), Ministero dell'Università e della Ricerca (NextGenEU 153D23001490006 M4C2.1.1, NextGenEU I53D23000820006 M4C2.1.1, NextGenEU I53D23001490006 M4C2.1.1, SOE2024\_0000023); Japan: Japan Society for the Promotion of Science (JSPS KAKENHI JP22H01227, JSPS KAKENHI JP22H04944, JSPS KAKENHI JP22KK0227, JSPS KAKENHI JP24K23939, JSPS KAKENHI JP24KK0251, JSPS KAKENHI JP25H00650, JSPS KAKENHI JP25H01291, JSPS KAKENHI JP25K01023); Norway: Research Council of Norway (RCN-314472); Poland: Ministry of Science and Higher Education (IDUB AGH, POB8, D4 no 9722), Polish National Science Centre (NCN 2021/42/E/ST2/00350, NCN OPUS 2023/51/B/ST2/02507, NCN OPUS nr 2022/47/B/ST2/03059, NCN UMO-2019/34/E/ST2/00393, UMO-2022/47/O/ST2/00148, UMO-2023/49/B/ST2/04085, UMO-2023/51/B/ST2/00920, UMO-2024/53/N/ST2/00869); Portugal: Foundation for Science and Technology (FCT); Spain: Agencia de Gestión de Ayudas Universitarias y de Investigación (AGAUR - 2023 BP 00141), Generalitat Valenciana (ASFAE/2022/008), Ministry of Science and Innovation (MCIN \& NextGenEU PCI2022-135018-2, MICIN \& FEDER PID2021-125273NB, RYC2019-028510-I, RYC2020-030254-I, RYC2021-031273-I, RYC2022-038164-I), Ministerio de Ciencia, Innovación y Universidades/Agencia Estatal de Investigaci\'on (PID2022-142604OB-C22); Sweden: Carl Trygger Foundation (Carl Trygger Foundation CTS 22:2312), Swedish Research Council (Swedish Research Council 2023-04654, VR 2021-03651, VR 2022-03845, VR 2022-04683, VR 2023-03403, VR 2024-05451), Knut and Alice Wallenberg Foundation (KAW 2018.0458, KAW 2022.0358, KAW 2023.0366); Switzerland: Swiss National Science Foundation (SNSF - PCEFP2\_194658); United Kingdom: Royal Society (NIF-R1-231091); United States of America: U.S. Department of Energy (ECA DE-AC02-76SF00515), Neubauer Family Foundation.

%
%


%
\clearpage
\printbibliography
\clearpage
 
\begin{flushleft}
\hypersetup{urlcolor=black}
{\Large The ATLAS Collaboration}

\bigskip

\AtlasOrcid[0000-0002-6665-4934]{G.~Aad}$^\textrm{\scriptsize 104}$,
\AtlasOrcid[0000-0001-7616-1554]{E.~Aakvaag}$^\textrm{\scriptsize 17}$,
\AtlasOrcid[0000-0002-5888-2734]{B.~Abbott}$^\textrm{\scriptsize 123}$,
\AtlasOrcid[0000-0002-0287-5869]{S.~Abdelhameed}$^\textrm{\scriptsize 119a}$,
\AtlasOrcid[0000-0002-1002-1652]{K.~Abeling}$^\textrm{\scriptsize 55}$,
\AtlasOrcid[0000-0001-5763-2760]{N.J.~Abicht}$^\textrm{\scriptsize 49}$,
\AtlasOrcid[0000-0002-8496-9294]{S.H.~Abidi}$^\textrm{\scriptsize 30}$,
\AtlasOrcid[0009-0003-6578-220X]{M.~Aboelela}$^\textrm{\scriptsize 45}$,
\AtlasOrcid[0000-0002-9987-2292]{A.~Aboulhorma}$^\textrm{\scriptsize 36e}$,
\AtlasOrcid[0000-0001-5329-6640]{H.~Abramowicz}$^\textrm{\scriptsize 157}$
% ...
\bigskip
\\

% ...
$^{17}$Department for Physics and Technology, University of Bergen, Bergen; Norway.\\
% ...
$^{30}$Physics Department, Brookhaven National Laboratory, Upton NY; United States of America.\\
% ...
$^{36}$$^{(a)}$Facult\'e des Sciences Ain Chock, Universit\'e Hassan II de Casablanca;$^{(b)}$Facult\'{e} des Sciences, Universit\'{e} Ibn-Tofail, K\'{e}nitra;$^{(c)}$Facult\'e des Sciences Semlalia, Universit\'e Cadi Ayyad, LPHEA-Marrakech;$^{(d)}$LPMR, Facult\'e des Sciences, Universit\'e Mohamed Premier, Oujda;$^{(e)}$Facult\'e des sciences, Universit\'e Mohammed V, Rabat;$^{(f)}$Institute of Applied Physics, Mohammed VI Polytechnic University, Ben Guerir; Morocco.\\
% ...
$^{45}$Physics Department, Southern Methodist University, Dallas TX; United States of America.\\
% ...
$^{49}$Fakult\"{a}t Physik, Technische Universit{\"a}t Dortmund, Dortmund; Germany.\\
% ...
$^{55}$II. Physikalisches Institut, Georg-August-Universit\"{a}t G\"ottingen, G\"ottingen; Germany.\\
% ...
$^{104}$CPPM, Aix-Marseille Universit\'e, CNRS/IN2P3, Marseille; France.\\
% ...
$^{119}$$^{(a)}$New York University Abu Dhabi, Abu Dhabi;$^{(b)}$United Arab Emirates University, Al Ain; United Arab Emirates.\\
% ...
$^{123}$Homer L. Dodge Department of Physics and Astronomy, University of Oklahoma, Norman OK; United States of America.\\
% ...
$^{157}$Raymond and Beverly Sackler School of Physics and Astronomy, Tel Aviv University, Tel Aviv; Israel.\\
% ...
$^{a}$ Also at Affiliated with an institute formerly covered by a cooperation agreement with CERN.\\
$^{b}$ Also at An-Najah National University, Nablus; Palestine.\\
$^{c}$ Also at Borough of Manhattan Community College, City University of New York, New York NY; United States of America.\\
$^{d}$ Also at Center for Interdisciplinary Research and Innovation (CIRI-AUTH), Thessaloniki; Greece.\\
$^{e}$ Also at Centre of Physics of the Universities of Minho and Porto (CF-UM-UP); Portugal.\\
$^{f}$ Also at CERN, Geneva; Switzerland.\\
% ...

\end{flushleft}


%

%
%

%

%
%
%
%

\end{document}